\documentclass[preprint,3p,12]{elsarticle}%
\usepackage{amsfonts}
\usepackage{amsmath}
\usepackage{amssymb}
\usepackage{algorithm}
\usepackage{algorithmicx}
\usepackage{algpseudocode}
\usepackage{amsmath}
\usepackage{graphicx}
\usepackage{setspace}
\usepackage{ntheorem}%
\providecommand{\U}[1]{\protect\rule{.1in}{.1in}}
\providecommand{\U}[1]{\protect\rule{.1in}{.1in}}
\providecommand{\U}[1]{\protect\rule{.1in}{.1in}}
\providecommand{\U}[1]{\protect\rule{.1in}{.1in}}
\providecommand{\U}[1]{\protect\rule{.1in}{.1in}}
\providecommand{\U}[1]{\protect\rule{.1in}{.1in}}
\providecommand{\U}[1]{\protect\rule{.1in}{.1in}}
\newtheorem*{theorem}{Theorem}

\journal{Signal Processing}
\begin{document}

\begin{frontmatter}
\title{On the Reconstruction of Randomly Sampled Sparse Signals Using an Adaptive Gradient Algorithm}
\author{Ljubi\v{s}a Stankovi\'{c}, Milo\v{s} Dakovi\'{c}}
\address{University of Montenegro, Podgorica, ljubisa@ac.me, milos@ac.me}
\begin{abstract}
Sparse signals can be recovered from a reduced set of samples by using
compressive sensing algorithms. In common methods the signal is recovered in the sparse domain.
A method for the reconstruction of sparse signal which reconstructs the remaining
missing samples/measurements is recently proposed.
The available samples are fixed, while the missing samples are considered as
minimization variables. Recovery of missing samples/measurements is
done using an adaptive  gradient-based algorithm in the time domain. A new
criterion for the parameter adaptation in this algorithm, based on the
gradient direction angles, is proposed. It improves the algorithm
computational efficiency. A theorem for the uniqueness of the recovered signal for given set of missing samples
(reconstruction variables) is presented. The case when available samples are a random
subset of a uniformly or nonuniformly sampled signal is considered in this paper. A
recalculation procedure is used to reconstruct the nonuniformly sampled signal.
The methods are illustrated on statistical examples.
\end{abstract}
\end{frontmatter}

\section{Introduction}

A discrete-time signal can be transformed into other domains in various ways.
Some signals that cover whole considered interval in one domain could be
sparse in a transformation domain. Compressive sensing theory, in general,
deals with a lower dimensional set of linear observations of a sparse signal,
in order to recover all signal values \cite{donoho2006}-\cite{GradientIET}.
This area intensively develops in the last decade. A common form of
observations are the signal samples. A reduced set of samples can be
considered within the compressive sensing theory in order to represent a
signal with the lowest possible number of samples. This theory may be applied
to the situations when the missing samples are not a result of compressive
sensing strategy, with the aim to reduce data size. In many engineering
applications the signal samples are missing due to the system physical
constraints or unavailability of the measurements. It can also happen that
some arbitrarily positioned samples of a signal are so heavily corrupted by
disturbances that it is better to omit them and consider as missing in the
analysis \cite{Imp}, \cite{Nas}. It is interesting to note that one of the
first compressive sensing theory successes in applications (computed
tomography reconstruction) was not related to the intentional compressive
sensing strategy but to the physical problem constraints, restricting the set
and positions of the available data. Signal reconstruction, with arbitrary
missing samples, not being a result of any intentional compressive sensing
strategy, is the topic of this paper.

Several approaches to reconstruct sparse signals from their random lower
dimensional set of linear observations are introduced \cite{Gradient1}%
-\cite{daubechies2004}. The most common are the reconstruction algorithms
based either on the gradient formulations \cite{Gradient1} or the orthogonal
matching pursuit approaches \cite{mallat1993}.

Recently a method for the reconstruction of a sparse signal with
disturbed/missing samples has been proposed \cite{GradientIET}. In contrast to
the common reconstruction methods that recover the signal in their sparsity
domain the proposed method reconstructs missing samples/measurements to make
the set of samples/measurements complete. Since the available samples are
fixed, the minimization variables are the missing samples. It means that the
number of variables is equal to the number of missing signal samples in the
observation domain. A new criterion for the parameter adaptation of this
simple algorithm, based on the gradient directions, is proposed in this paper.
Computational time to achieve the target reconstruction accuracy is
significantly reduced with respect to the original criterion in
\cite{GradientIET}. The Fourier transform domain is used as a case study,
although the algorithm application is not restricted to this transform
\cite{IS}. The algorithm efficiency is statistically checked. After the
recovery of sparse signal, then its uniqueness is checked by using the proposed theorem. Its application is quite simple and numerically efficient.

Sparse signals with available samples being a random subset of nonuniformly
sampled signal, not corresponding to the uniform sampling grid, are considered
as well \cite{D1}-\cite{D5}. This case belongs to class of signals with
indirect measurements. A possibility to recalculate the signal values at the
sampling theorem positions is exploited in the nonuniform case before a
reconstruction algorithm is applied \cite{NUS}.

The paper is organized as follows. After the definitions in the next section,
the adaptive gradient algorithm, with a new criterion for the algorithm
parameter adaptation, is presented in Section 3. Uniqueness of the obtained
solution is analyzed in Section 3. Reconstruction of nonuniformly sampled
sparse signals is considered in Section 4.

\section{Gradient-based Reconstruction}

Consider a discrete-time signal $x(n)$ with $N$ samples whose transform
coefficients are $X(k)=T[x(n)]$. Signal $x(n)$ is sparse in the transformation
domain if the number of nonzero transform coefficients $s$ is much lower than
the number of the original signal samples $N$, $s\ll N$, i.e., $X(k)=0$ for
$k\notin\{k_{1}$, $k_{2}$, ..., $k_{s}\}$. The DFT will be considered in this
paper, when $X(k)=\mathrm{DFT}[x(n)].$ \ \ Assume that a set of $M<N$ signal
samples in the time domain is available at the instants corresponding to the
discrete-time positions
\begin{equation}
n_{i}\in\mathbb{N}_{A}=\{n_{1},n_{2},...,n_{M}\}\subset\mathbb{N}%
=\{0,1,2,...,N-1\}. \label{Set_AV}%
\end{equation}
In general, the signal recovery within the compressive sensing framework
consists in reconstructing the signal (calculation of
missing/unavailable/discarded samples) so that the number of nonzero transform
coefficients $X(k)$ is minimal, subject to the available sample values.
Counting of the nonzero transform coefficients is achieved by a simple
mathematical form $\left\Vert X(k)\right\Vert _{0}$, sometimes referred to as
the \textquotedblleft$\ell_{0}$-norm\textquotedblright, \cite{donoho2006},
\cite{Ljubisa}. Thus, the problem statement is
\begin{equation}
\min\left\Vert X(k)\right\Vert _{0}\text{ subject to }\mathbf{y=AX},
\label{DEF_L0}%
\end{equation}
where $\mathbf{y}=[x(n_{1})~~x(n_{2})~~...~~x(n_{M})]^{T}$
is the vector of available signal samples, $\mathbf{X}=[X(1)~~X(2)~~...~~X(N)]^{T}$ is the vector of unknown
transform coefficients, and $\mathbf{A}$ is the inverse
transform matrix with omitted rows corresponding to the unavailable signal
samples. The $\ell_{0}$-norm based formulation is an NP-hard combinatorial
optimization problem. Its calculation complexity is of order $\binom{N}{s}$.
In theory, the NP-hard problems can be solved by an exhaustive search.
However, as the problem parameters $N$ and $s$ increase the running time
increases and the problem becomes unsolvable. These are the reasons why the
$\ell_{1}$-norm of the signal transform, $\left\Vert X(k)\right\Vert _{1}%
=\sum_{k=0}^{N-1}\left\vert X(k)\right\vert $, is commonly used as a sparsity
measure function. The minimization problem is
\begin{equation}
\min\sum_{k=0}^{N-1}\left\vert X(k)\right\vert \text{ subject to
}\mathbf{y=AX.} \label{DEF_L1}%
\end{equation}
This minimization problem, under the conditions defined within the restricted
isometry property (RIP), \cite{RIP}, \cite{cosamp}, can produce the same
result as (\ref{DEF_L0}). \ Note that other norms $\ell_{p}$\textit{ }between
the $\ell_{0}$-norm and the $\ell_{1}$-norm, with values $0<p<1$, are also
used in the minimization in attempts to combine good properties of these two
norms \cite{donoho2006,GradientIET,L095}.

\subsection{Algorithm}

A simple gradient-based algorithm that iteratively calculates the missing
sample values according to (\ref{DEF_L1}), is presented next. The basic idea
for the algorithm comes from the gradient-based minimization approaches. The
missing samples are considered as variables. The influence of their variations
on the sparsity measure is checked \cite{GradientIET}. By performing an
iterative procedure, the missing samples are changed toward lower sparsity
measure values, in order to approach the minimum of the convex $\ell_{1}$-norm
based sparsity measure (\ref{DEF_L1}). If the recovery conditions for the
$\ell_{1}$-norm \cite{RIP} are met then the $\ell_{1}$-norm minimum will be at
the same position as the $\ell_{0}$-norm minimum, representing the true values
of the missing samples.

\subsection{Review of the Algorithm}

The initial signal $y^{(0)}(n)$ is defined for $n\in$ $\mathbb{N}%
=\{0,1,\ldots,N-1\}$ as:
\begin{equation}
y^{(0)}(n)=\left\{
\begin{array}
[c]{ll}%
0 & \text{ for missing samples, }n\in\mathbb{N}_{Q}\\
x(n) & \text{ for available samples, }n\in\mathbb{N}_{A}%
\end{array}
\right.  , \label{initSig}%
\end{equation}
where $\mathbb{N}_{Q}$ is the complement of $\mathbb{N}_{A}$ with respect to
$\mathbb{N}$ defined by (\ref{Set_AV}).

The missing signal samples are then corrected in an iterative procedure as
\begin{equation}
y^{(m)}(n)=y^{(m-1)}(n)-g^{(m)}(n),
\end{equation}
where $g^{(m)}(n)$ is an estimate of the sparsity measure gradient vector
coordinate along the variable $y(n)$ direction, in the $m$th iteration. At the
positions of the available signal samples, $n\in\mathbb{N}_{A}$,
$g^{(m)}(n)=0$. At the positions of missing samples, $n_{i}\in\mathbb{{N}}%
_{Q}$, its values are calculated by changing the signal values and forming new
signals $y_{1}(n)$ and $y_{2}(n)$ as
\begin{align}
y_{1}(n)  &  =y^{(m)}(n)+\Delta\delta(n-n_{i})\nonumber\\
y_{2}(n)  &  =y^{(m)}(n)-\Delta\delta(n-n_{i}). \label{Sig_Delta}%
\end{align}
The algorithm step is denoted by $\Delta$. The values of $g^{(m)}(n_{i})$ at
$n_{i}\in\mathbb{N}_{Q}$ are%
\begin{equation}
g(n_{i})=\frac{\sum_{k=0}^{N-1}\left\vert Y_{1}(k)\right\vert -\sum
_{k=0}^{N-1}\left\vert Y_{2}(k)\right\vert }{N}, \label{eq:mjera}%
\end{equation}
where $Y_{1}(k)=\mathrm{DFT}[y_{1}(n)]$ and $Y_{2}(k)=\mathrm{DFT}[y_{2}(n)]$.

The initial value for the algorithm adaptation step $\Delta$ is estimated as%
\begin{equation}
\Delta=\max_{n\in\mathbb{N}_{A}}|x(n)|=\max_{n}\left\vert y^{(0)}%
(n)\right\vert . \label{deltaINIt}%
\end{equation}

The gradient algorithm will approach the minimum point of the $\ell_{1}$-norm
based sparsity measure with a precision related to the algorithm step $\Delta$.

\subsubsection{Stopping Criterion and Adaptive Step}

Rate of the algorithm convergence for different steps is considered in
\cite{GradientIET}. The algorithm performance is significantly improved by
using adaptive step $\Delta$. A criterion that efficiently detects the event
that the algorithm has reached the vicinity of the sparsity measure minimum is
proposed in this paper. It is based on the direction change of the gradient
vector. When the vicinity of the optimal point is reached, the gradient
estimate in the $\ell_{1}$-norm based sparsity measure function changes
direction for almost $180$ degrees. For each two successive gradient
estimations $g^{(m-1)}(n)$ and $g^{(m)}(n)$, the angle $\beta_{m}$ between
gradient vectors is calculated as%
\[
\beta_{m}=\arccos\frac{\sum_{n=0}^{N-1}g^{(m-1)}(n)g^{(m)}(n)}{\sqrt
{\sum_{n=0}^{N-1}\left(  g^{(m-1)}(n)\right)  ^{2}}\sqrt{\sum_{n=0}%
^{N-1}\left(  g^{(m)}(n)\right)  ^{2}}}.
\]
If the angle $\beta_{m}$ is above $170^{\circ}$ it means that the values
reached oscillatory nature around the minimal measure value position. When
this kind of the angle change is detected the step $\Delta$ is reduced, for
example, $\Delta/\sqrt{10}\rightarrow\Delta$, and the same calculation
procedure is continued from the reached reconstructed signal values. When the
optimal point is reached with a sufficiently small $\Delta$, then this value
of $\Delta$ is also an indicator of the solution precision. Value of $\Delta$
can be used as the algorithm stopping criterion.

A common way to estimate the precision of the result in iterative algorithms
is based on the change of the result in the last iteration. An average of
changes in last iteration in a large number of missing samples is a good
estimate of the achieved precision. Thus, the value of
\[
T_{r}=10\log_{10}\frac{\sum_{n\in\mathbb{N}_{Q}}|y_{p}(n)-y^{(m)}(n)|^{2}%
}{\sum_{n\in\mathbb{N}_{Q}}|y^{(m)}(n)|^{2}}.
\]
can be used as an rough estimate of the reconstruction error to signal ratio.
Here $y_{p}(n)$ is the reconstructed signal prior to $\Delta$ reduction (prior
to the execution of the algorithm inner loop, lines 7-20 in Algorithm 1) and
$y^{(m)}(n)$ is the reconstructed signal after the inner loop execution. This
value can also be used as a criterion to stop the algorithm. If $T_{r}$ is
above the required precision threshold $T_{max}$ (for example, if
$T_{r}>-100dB$), the calculation procedure should be repeated with smaller
values $\Delta$.

A pseudo code of this algorithm is presented in Algorithm 1.

\algnewcommand\algorithmicoutput{\textbf{Output:}} \algnewcommand\Output{\item[\algorithmicoutput]}

\begin{algorithm}
\setstretch{1.1}
\caption{Reconstruction}
\label{RecAlg}
\begin{algorithmic}[1]
\Require
\Statex
\begin{itemize}
\item Set of missing/omitted sample positions $\mathbb{N}_{Q}$
\item Available samples $x(n)$, $n\notin \mathbb{N}_{Q}$
\end{itemize}
\Statex
\State \label{alg:init1}Set $y^{(0)}(n) \gets x(n)$ \Comment{for $n\notin \mathbb{N}_{Q}$}
\State \label{alg:init2}Set $y^{(0)}(n) \gets 0$ \Comment{for $n\in \mathbb{N}_{Q}$}
\State \label{alg:init3}Set $m \gets 0$
\State \label{alg:init4}Set $\Delta \gets \max{|y^{(0)}(n)|}$
\Repeat \label{alg:loop1}
\State Set $y_p(n)=y^{(m)}(n)$ \Comment{for each $n$}
\Repeat \label{alg:loop2}
\State $m \gets m+1$
\For{ $n_i\gets 0$ to $N-1$} \label{alg:for}
\If{ $n_i\in \mathbb{N}_{Q}$}
\State \label{alg:dft1}$Y_1(k) \gets\operatorname*{DFT}\{y^{(m)}(n)+\Delta \delta(n-n_i)\}$
\State \label{alg:dft2}$Y_2(k) \gets \operatorname*{DFT}\{y^{(m)}(n)-\Delta \delta(n-n_i)\}$
\smallskip
\State \label{alg:grad}$\displaystyle g^{(m)}(n_i) \gets \frac{1}{N}{\sum_{k=0}^{N-1}|Y_1(k)|-|Y_2(k)|}$
\Else
\State $  g^{(m)}(n_i) \gets 0$
\EndIf
\State \label{alg:adjust}$y^{(m+1)}(n_i) \gets y^{(m)}(n_i)-  g^{(m)}(n_i)$
\EndFor \label{alg:next}
\smallskip
\State $\displaystyle \beta_m=\arccos \frac{\sum_{n=0}^{N-1}g^{(m-1)}(n) g^{(m)}(n)}{\sqrt{\sum_{n=0}^{N-1}\left( g^{(m-1)}(n) \right)^2}  \sqrt{\sum_{n=0}^{N-1}\left( g^{(m)}(n) \right)^2}}$
\medskip
\Until{ $\beta_m<170^{\circ}$}
\State $\Delta \gets \Delta/\sqrt{10}$ \label{alg:delta3}
\smallskip
\State $\displaystyle T_{r}=10\log_{10}\frac{\sum_{n\in \mathbb{N}_{Q}}|y_{p}(n)-y^{(m)}%
(n)|^{2}}{\sum_{n\in \mathbb{N}_{Q}}|y^{(m)}(n)|^{2}}.
$
\smallskip
\Until{ $T_r<T_{max} $}
\State \Return $y^{(m)}(n)$
\Statex
\Output
\Statex
\begin{itemize}
\item Reconstructed signal $x_R(n)=y^{(m)}(n)$
\end{itemize}
\end{algorithmic}
\end{algorithm}

\textbf{Comments on the algorithm:}

- The inputs to the algorithm are the signal length $N$, the set of available
samples $\mathbb{N}_{A}$, the available signal values $x(n_{i})$, $n_{i}\in$
$\mathbb{N}_{A}$, and the required precision $T_{max}$.

- Instead of calculating signals (\ref{Sig_Delta}) and their $\mathrm{DFTs}$
for each $n_{i}\in\mathbb{N}_{Q}$ we can calculate
\begin{align*}
\left\vert Y_{1}(k)\right\vert  &  =\left\vert Y^{(m)}(k)+\Delta D_{n_{i}%
}(k)\right\vert \\
\left\vert Y_{2}(k)\right\vert  &  =\left\vert Y^{(m)}(k)-\Delta D_{n_{i}%
}(k)\right\vert
\end{align*}
with $Y^{(m)}(k)=\mathrm{DFT}[y^{(m)}(n)]$ and $D_{n_{i}}(k)=\mathrm{DFT}%
[\delta(n-n_{i})]=\exp(-j2\pi n_{i}k/N)$, for each $n_{i}\in\mathbb{N}_{Q}$.
Since $D_{n_{i}}(k)$ are independent of the iteration number $m$ they can be
calculated only once, independently from the DFT of the signal.

- In a gradient-based algorithm, a possible divergence is related to the
algorithm behavior for a large step $\Delta$. Small steps influence the rate
of the algorithm approach to the solution only, with the assumption that it
exists. Influence of small steps to the calculation complexity is considered
in \cite{GradientIET}. Here, we will examine the algorithm behavior for a
large value of step $\Delta$. We can write
\begin{align*}
\left\vert Y_{1}(k)\right\vert -\left\vert Y_{2}(k)\right\vert  &  =\left\vert
Y^{(m)}(k)+\Delta D_{n_{i}}(k)\right\vert -\left\vert Y^{(m)}(k)-\Delta
D_{n_{i}}(k)\right\vert \\
&  =\Delta\left\vert D_{n_{i}}(k)\right\vert \left(  \left\vert 1+\frac
{Y^{(m)}(k)}{\Delta D_{n_{i}}(k)}\right\vert -\left\vert 1-\frac{Y^{(m)}%
(k)}{\Delta D_{n_{i}}(k)}\right\vert \right)  .
\end{align*}
Considering the complex number $a=Y^{(m)}(k)/(\Delta D_{n_{i}}(k)),$ with
$|a|\ll1$ for a large $\Delta$, from the problem geometry it is easy to show
that the following bounds hold $0\leq\left\vert |1+a|-|1-a|\right\vert
\leq2\left\vert a\right\vert $. Therefore,%
\[
0\leq\Big\vert\left\vert Y_{1}(k)\right\vert -\left\vert Y_{2}(k)\right\vert
\Big\vert\leq2\left\vert Y^{(m)}(k)\right\vert .
\]
Lower limit $0$ is obtained if $a$ is imaginary-valued, while the upper limit
$2\left\vert Y^{(m)}(k)\right\vert $ follows if $a$ is real-valued.

It means that the value of the finite difference $\left\vert Y_{1}%
(k)\right\vert -\left\vert Y_{2}(k)\right\vert ,$ that is used to correct the
missing signal samples, does not depend on the value of the step $\Delta$, if
$\Delta$ is large. The missing signal values will be adapted for a value
independent on $\Delta$ in that case. The values of missing samples will
oscillate within the range of the original signal values of order $\left\vert
Y^{(m)}(k)\right\vert /N$, until $\Delta$ is reduced in the iterations. Then
the missing samples will start approaching to the position of the sparsity
measure minimum. The initial values will be arbitrary changed within the
signal amplitude order as far as $\Delta$ is too large. It will not influence
further convergence of the algorithm, when the step $\Delta$ assumes
appropriate values.

- Since two successive gradient vectors are required to calculate the gradient
angle $\beta_{m}$, it is calculated starting from the second iteration for
each $\Delta$.

- The algorithm output is the reconstructed signal $y(n)$, $n=0,1,\ldots,N-1$.

- Other signal transforms can be used instead of the DFT. The only requirement
is that signal is sparse in that transform domain.

\begin{figure}
[ptb]
\begin{center}
\includegraphics[
height=1.8447in,
width=6.6919in
]%
{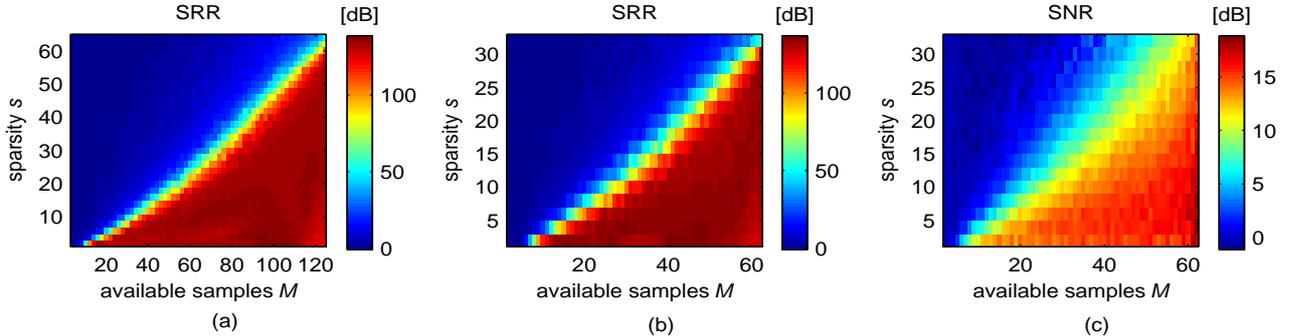}%
\caption{Signal-to-reconstruction-error (SRR) obtained by using Algorithm
\ref{RecAlg}, averaged over 100 realizations for various sparsity $s$ and
number of available samples $M$: (a) The total number of samples is $N=128$.
(b) The total number of samples is $N=64.$ (c) With a Gaussian noise in the
input signal, $SNR=20$ [dB] and $N=64.$}%
\label{fig1_fin_crtanje_b}%
\end{center}
\end{figure}

\textit{Example 1:} Consider a signal
\begin{equation}
x(t)=\sum_{i=1}^{K}A_{i}\cos(2\pi tk_{i}/T+\varphi_{i}),\label{signal}%
\end{equation}
with $t=n\Delta t,$ $\Delta t=1$, and the total number of samples $N=T/\Delta
t$. The sparsity parameter $s=2K$ is changed from $s=2$ to $s=N/2$. The
amplitudes $A_{i}$, frequencies $k_{i}$, and phases $\varphi_{i}$ are taken
randomly. Amplitude values are modeled as Gaussian random variables with
variance $1,$ the frequency indices assume random numbers within $1\leq
k_{i}\leq N-1$, and the phases assume uniform random values within $0\leq
\phi_{i}\leq2\pi$, in each realization. The reconstruction is performed by
using $100$ realizations for each $s$ with random sets of missing $N-M$
samples in each realization. The simulations are done for $N=128$ and for
$N=64$. The reconstructed signals $x_{R}(n)$ are obtained. The results are
presented in Fig.\ref{fig1_fin_crtanje_b}(a) and (b) in a form of the
signal-to-reconstruction-error ratio (SRR) in [dB]%
\begin{equation}
SRR=10\log\frac{\sum_{n=0}^{N-1}\left\vert x(n)\right\vert ^{2}}{\sum
_{n=0}^{N-1}\left\vert x(n)-x_{R}(n)\right\vert ^{2}}.\label{SRRR}%
\end{equation}
%

It is important to note that the all signal samples are used in this error
calculation. It means that possible nonunique solutions, satisfying the same
set of available samples, are considered as the reconstructions with
significant error since they significantly differ at the positions of
missing/recovered samples. This kind of reconstruction and uniqueness analysis
is exact. However, it can be used in simulations only since it requires the
exact signal in all samples. A check of the solution uniqueness, using the
missing sample positions only, will be discussed later.

Red colors indicate the region where the algorithm had fully recovered missing
samples (compared to the original samples) in all realizations, while blue
colors indicate the region where the algorithm could not recover missing
samples in any realization. In the transition region for $M$ slightly greater
than $2s$ we have cases when the signal recovery is not achieved and the cases
of full signal recovery.  A stopping criterion for the accuracy of $120$ [dB]
is used. It corresponds to a precision in the recovered signal of the same
order as in input samples, if they are acquired by a 20-bit A/D converter. The
case with $N=64$ is repeated with an additive input Gaussian noise such that
the input signal-to-noise ratio is $20$ [dB] in each realization
Fig.\ref{fig1_fin_crtanje_b}(c). The reconstruction error in this case is
limited by the input signal-to-noise value.

\begin{figure}[ptb]
\begin{center}
\includegraphics[
width=5.5359in
]{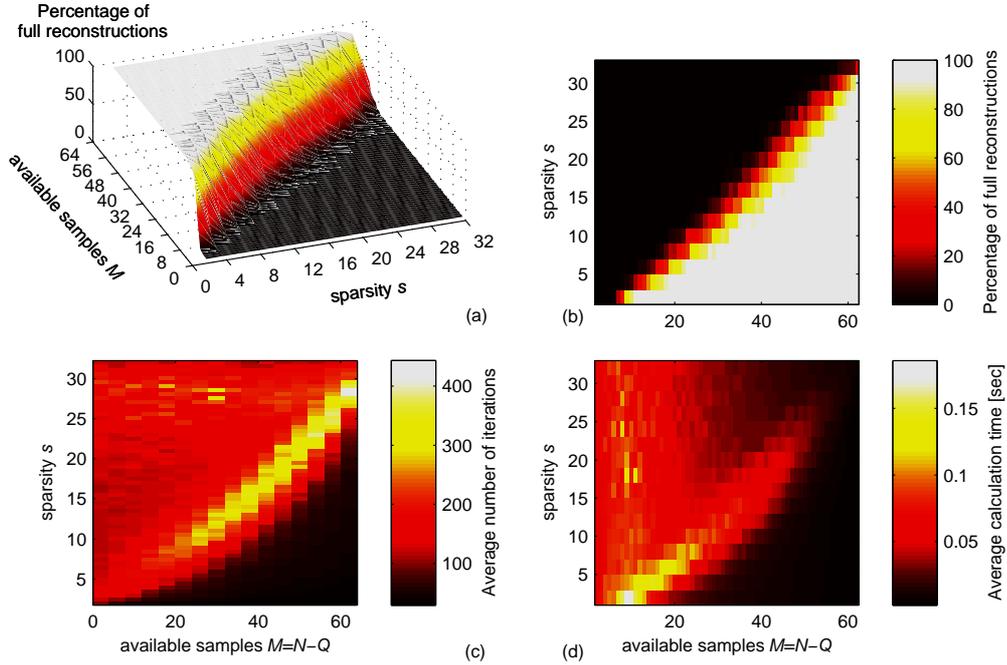}
\end{center}
\caption{(a)-(b) The percentage of the full recovery events as a function of
the number of available samples $M$ and the sparsity $s$ in the case of
$N=64$. (c) The average number of iterations as a function of the number of
missing samples and sparsity. (d) The average computation time. }%
\label{graphics_gradrec_mse_komp_3gr_4s}%
\end{figure}

The average reconstruction error in the noise-free cases is related to the
number of the full recovery events. For $N=64$ the number of the full recovery
events is checked and presented in Fig.\ref{graphics_gradrec_mse_komp_3gr_4s}
(a),(b). The average number of the algorithm iterations to produce the
required precision, as a function of the number of missing samples and signal
sparsity $s$, is presented as well, Fig.\ref{graphics_gradrec_mse_komp_3gr_4s}%
(c), along with the corresponding average computation time (in seconds) for
the Windows PC with Intel Dual Core processor,
Fig.\ref{graphics_gradrec_mse_komp_3gr_4s}(d). The average computation time is
proportional to the average number of iterations multiplied by the number of
missing samples (variables) $Q=N-M$.

Finite difference method and the adaptation procedure presented in this paper
overcome the problem of the derivative existence in the case of the $\ell_{1}%
$-norm near the optimal point. Although the main point of this manuscript is
to present a new method of reconstruction, with missing samples being
minimization variables, efficiency of the presented algorithm is compared with
the standard routines where the $\ell_{1}$-norm problem is solved using the
linear programming. Direct adaptation of missing samples can be used in
various applications (including recovery of sampled signals where a linear
relation between signal and transform cannot be established). The performance
of the proposed algorithm are compared with the algorithm that recasts the
recovery problem (6) into a linear programming framework and uses the primal-dual
interior point method (L1-magic code in MATLAB). Both algorithms are run with
the default parameters using 100 sparse signals with random parameters. The
results are presented in the Table \ref{Table_1}. Columns notation in the
table is: $s$ for sparsity, $Q=N-M$ for the number of missing samples, MAE
stands for the mean absolute error, LP-DP denotes the values obtained by
running the linear programming primal-dual algorithm (MATLAB L1-magic code),
and AS is for the presented adaptive algorithm with variable step. Calculation
time using MATLAB is presented in both cases.

\begin{table}[ptb]
\caption{MAE and elapsed time for L1-magic (LP-DP) and proposed algorithm
(AS)}%
\label{Table_1}
\centering \setstretch{1.35}
\begin{tabular}
[c]{cccccc}%
$s$ & $N-M$ & MAE LP-DP & MAE AS & time LP-DP & time AS\\\hline
6 & 16 & $1.719\times10^{-4}$ & $3.959\times10^{-7}$ & 0.043390 & 0.013433\\
10 & 16 & $1.124\times10^{-4}$ & $3.730\times10^{-7}$ & 0.041121 & 0.013393\\
16 & 16 & $2.575\times10^{-4}$ & $5.943\times10^{-7}$ & 0.041569 & 0.014003\\
6 & 32 & $3.238\times10^{-4}$ & $8.000\times10^{-7}$ & 0.038492 & 0.025733\\
10 & 32 & $3.454\times10^{-4}$ & $1.133\times10^{-6}$ & 0.038578 & 0.027270\\
16 & 32 & $1.068\times10^{-3}$ & $1.818\times10^{-6}$ & 0.046595 & 0.029636\\
6 & 45 & $1.000\times10^{-3}$ & $1.295\times10^{-6}$ & 0.041317 & 0.036442\\
10 & 45 & $4.731\times10^{-3}$ & $1.878\times10^{-6}$ & 0.039162 & 0.041843\\
16 & 45 & $2.415\times10^{-3}$ & $2.751\times10^{-6}$ & 0.042910 &
0.054350\\\hline
\end{tabular}
\end{table}

\begin{figure}
[ptb]
\begin{center}
\includegraphics[
height=2.8609in,
width=3.3184in
]%
{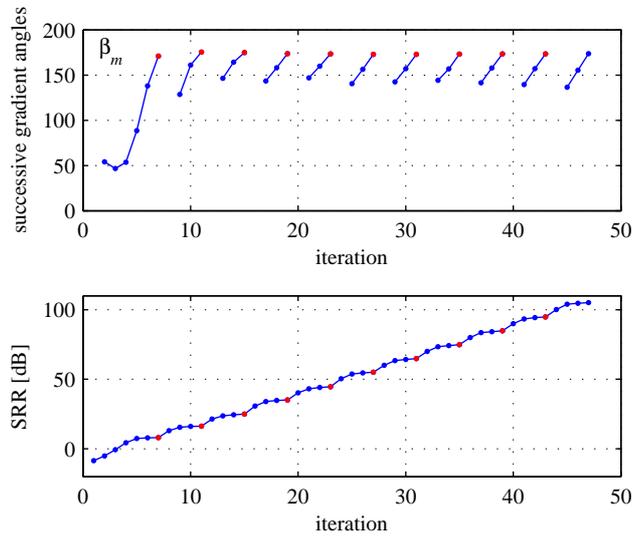}%
\caption{Angle between successive gradient estimations $\beta_{m}$ and the
signal-to-reconstruction-error ratio (SRR) as a function of the number of
iterations in the algorithm for one signal realization with 6 nonzero DFT
coefficients and $M=64$. }%
\label{mse_angles_gr}%
\end{center}
\end{figure}

An illustration of the algorithm performance regarding to the SRR and the
gradient angle $\beta_{m}$ in one realization, with $s=6$, is presented in
Fig.\ref{mse_angles_gr}.%

\section{Uniqueness of the Obtained Solution}

Uniqueness of the solution is guarantied if the restricted isometry property
is used and checked, with appropriate isometry constant for the norm-one based
minimization. However, two problem exist in the implementation of this method.
For a specific measurement matrix it produces quite conservative bounds,
meaning in practice a large number of false alarms for nonuniqueness. In
addition the check of the restricted isometry property required a
combinatorial approach, which is an NP hard problem (like the solution of the
problem using zero-norm in minimization). The the gradient-based algorithm
presented here considers missing samples/measurements as the minimization
variables. A theorem for the solution uniqueness, in terms of the missing
sample positions is used here. The proof of this theorem, with additional
details and examples, is given in \cite{Uniq}. When the reconstruction of a
signal is done and the solution of sparsity $s$ in the DFT domain is obtained,
then the theorem provides an easy check for the solution uniqueness.

Consider a signal
\[
x_{a}(n)=x(n)+z(n)
\]
where $z(n)=0$ for $n\in\mathbb{M}$ and takes arbitrary values at the
positions of missing samples $n=q_{m}\in\mathbb{N}_{Q}=\{q_{1},q_{2}%
,....,q_{Q}\}$. The DFT of this signal is
\begin{align*}
X_{a}(k)  &  =X(k)+Z(k)\\
&  =\sum_{i=1}^{s}\sigma_{i}\delta(k-k_{0i})+\sum_{m=1}^{Q}z(q_{m})e^{-j2\pi
q_{m}k/N}.
\end{align*}
The values of missing samples $y(n)=x_{a}(n)$ for $n\in\mathbb{N}_{Q}$ are
considered as variables, in the sparsity measure minimization process, with
the final goal to get $x_{a}(n)=x(n)$, or $z(n)=0$ for all $n$. Existence of
unique solution depends on the number of missing samples $Q$, their positions
$\mathbb{N}_{Q}$, and the available signal values \cite{Uniq}. The uniqueness
here means that if a sparse signal, with the transform $X(k)$, is
reconstructed using the fixed set of available samples and the gradient-based
algorithm (or any other reconstruction algorithm), then there is no other
signal transform of the same or lower sparsity satisfying the same set of
available sample values.

\begin{theorem}
Consider signal $x(n)$ that is sparse in the DFT domain with unknown sparsity.
Assume that the signal length is $N=2^{r}$ samples and that $Q$ samples are
missing at the instants $q_{m}\in\mathbb{N}_{Q}$. Also assume that the
reconstruction is performed and that the DFT of reconstructed signal is of
sparsity $s$. Assume that the positions of the reconstructed nonzero values in
the DFT are $k_{0i}\in\mathbb{K}_{s}=\{k_{01},k_{02},....,k_{0s}\}.$
Reconstruction result is unique if the inequality
\[
s<N-\max_{h=0,1,...,r-1}\left\{  2^{h}\left(  Q_{2^{h}}-1\right)
-s+2S_{2^{r-h}}\right\}
\]
holds. Integers $Q_{2^{h}}$ and $S_{2^{r-h}}$ are calculated as%
\begin{align*}
Q_{2^{h}}  &  =\max_{b=0,1,...,2^{h}-1}\{\operatorname{card}\{q:q\in
\mathbb{N}_{Q}\text{ and }\operatorname{mod}(q,2^{h})=b\}\}\\
S_{2^{r-h}}  &  =\sum_{l=1}^{Q_{2^{h}}-1}P_{h}(l)\\
P_{h}(l)  &  =\underset{b=0,1,...,2^{r-h}-1}{\mathrm{sort}}%
\{\operatorname{card}\{k:k\in\mathbb{K}_{s}\text{ and }\operatorname{mod}%
(k,2^{r-h})=b\}\}
\end{align*}
where $P_{h}(1)\leq P_{h}(2)\leq...\leq P_{h}(2^{r-h})$.

Signal independent uniqueness corresponds to the worst case signal form, when
$S_{2^{r-h}}=0$.
\end{theorem}

The answer is obtained almost immediately, since the computational complexity
of the Theorem is of order $O(N)$. The proof is given in \cite{Uniq}.

\section{Random Subset of Nonuniformly Sampled Values}

Consider now a discrete-time signal obtained by sampling a continuous-time
signal $x(t)$ at arbitrary positions. Since the DFT will be used in the
analysis, we can assume that the continuous-time signal is periodically
extended with a period $T$. According to the sampling theorem, the period $T$
is related to the number of samples $N$, the sampling interval $\Delta t$, and
the maximal frequency $\Omega_{m}$ as $\Omega_{m}=\pi/\Delta t=\pi N/T$. The
continuous-time signal can be written as an inverse Fourier series
\begin{equation}
x(t)=\sum_{k=-N/2}^{N/2-1}X_{k}e^{j2\pi kt/T},
\end{equation}
with the Fourier series coefficients being related to the DFT as
$X_{k}N=X(k)=\mathrm{DFT}[x(n)]$ and $x(n)=x(n\Delta t)$. The discrete-time
index $n$ corresponds to the continuous-time instant $t=n\Delta t$.
Discrete-frequency indices$\ $are $k\in\{-N/2,...,-1,0,1,...,N/2-1\}$. Any
signal value can be reconstructed from the samples taken according to the
sampling theorem,
\begin{equation}
x(t)=\sum_{n=0}^{N-1}x(n\Delta t)e^{j(n-t/\Delta t)\pi/N}\frac{\sin
[(n-\frac{t}{\Delta t})\pi]}{N\sin[(n-\frac{t}{\Delta t})\pi/N]}%
.\label{sig_rec_c}%
\end{equation}
This relation holds for an even $N$, \cite{NUS}, \cite{knjiga}.

A signal $x(t)$ is sparse in the transformation domain if the number of
nonzero transform coefficients $s$ is much lower than the number of the
original signal samples $N$ within $T$, $s\ll N$, i.e., $X_{k}=0$ for
$k\notin\{k_{1}$, $k_{2}$, ..., $k_{s}\}$. A signal%
\begin{equation}
x(t)=\sum_{k\in\{k_{1},k_{2},...,k_{s}\}}X_{k}e^{j2\pi kt/T}.
\label{SIG_FS_SAM}%
\end{equation}
of sparsity $s$ can be reconstructed from $M$ samples, where $M<N$, if the
recovery conditions are met.

Consider now a random set of possible sampling instants $\{t_{1}%
,t_{2},...,t_{N}\}$,
\[
t_{i}=i\Delta t+\nu_{i},
\]
where $\nu_{i}$ is a uniform random variable $-\Delta t/2\leq\nu_{i}\leq\Delta
t/2$. Here $t_{i}$ denotes a time instant, while in the uniform sampling the
discrete-time index $n_{i}$ has been used to indicate instant corresponding to
$n_{i}\Delta t$. Assume that a random set of $M$ signal samples is available
at
\[
t_{i}\in\mathbb{T}_{A}=\{t_{1},t_{2},...,t_{M}\}.
\]

Since the signal is available at randomly positioned instants the Fourier
transform coefficients estimated as
\[
\hat{X}_{k}=\sum_{t_{i}\in\mathbb{T}_{A}}x(t_{i})\exp(-j2\pi kt_{i}/T)
\]
will not be sparse even if a large number $M$ of samples is available. To
improve the results, the problem can be reformulated to produce a better
estimation of the sparse signal transform during the recovery process. If the
signal values were available at $t_{i}\in\mathbb{T}_{A}$ for $M=N$ the signal
values at the sampling theorem positions could be recovered. The
transformation matrix relating samples taken at $t_{i}$ with the signal values
at the sampling theorem positions, according to (\ref{sig_rec_c}), is
\begin{align*}
\left[
\begin{array}
[c]{c}%
x(t_{1})\\
x(t_{2})\\
...\\
x(t_{N})
\end{array}
\right]   &  =\left[
\begin{array}
[c]{cccc}%
b_{11} & b_{12} & ... & b_{1N}\\
b_{21} & b_{22} & ... & b_{2N}\\
... & ... & ... & ...\\
b_{N1} & b_{N2} & ... & b_{NN}%
\end{array}
\right]  \left[
\begin{array}
[c]{c}%
x(1)\\
x(2)\\
...\\
x(N)
\end{array}
\right] \\
\mathbf{\hat{x}}  &  \mathbf{=Bx}%
\end{align*}
with
\[
b_{ij}=\frac{\sin[(j-t_{i}/\Delta t)\pi]}{N\sin[(j-t_{i}/\Delta t)\pi
/N]}e^{j(j-t_{i}/\Delta t)\pi/N}%
\]
A problem here is that we know just $M<N$ of signal samples. The values at
unavailable positions $t_{i}\notin\mathbb{T}_{A}$ are assumed to be zero in
the initial iteration. Their positions are assumed at the sampling theorem
instants, $t_{i}=i\Delta t$ for $t_{i}\notin\mathbb{T}_{A}$, since they are
not known anyway. With this assumption the problem reduces to the missing
samples $q_{m}\in\mathbb{N}_{Q}=\{q_{1},q_{2},....,q_{Q}\}$, being
\ considered as variables and the remaining samples, defined by vector
$\mathbf{x}$, being calculated as
\begin{equation}
\mathbf{x=B}^{-1}\mathbf{\hat{x}}. \label{InV_SAM}%
\end{equation}
The matrix $\mathbf{B}^{-1}$ is inverted only once for the given signal sample
positions. There is a direct relation to calculate the values $x(n\Delta t)$
based on the randomly sampled values $x(t_{i})$, \cite{NUS}, where the
inversion is not needed.

The algorithm is adapted to this kind of signals as follows:

The signals $\hat{y}_{1}(t_{i})=y^{(m)}(t_{i})+\Delta\delta(t-t_{i})$ and
$\hat{y}_{2}(t_{i})=y^{(m)}(t_{i})-\Delta\delta(t-t_{i})$ are formed. The
available samples are recalculated to $y_{1}(n)$ and $y_{2}(n)$ according to
(\ref{InV_SAM}) as
\[
\mathbf{y}_{1}\mathbf{=B}^{-1}\mathbf{\hat{y}}_{1}%
\]
and
\[
\mathbf{y}_{2}\mathbf{=B}^{-1}\mathbf{\hat{y}}_{2}.
\]
These signals are used in the next algorithm steps.

\textit{Example 2:} Consider the signal defined by (\ref{signal}). Similar
results for the SRR and the average number of iterations, for various $M$ and
$s$, are obtained here as in Fig.\ref{fig1_fin_crtanje_b}. Thus, they will not
be repeated. Instead we will present a particular realization with $s=6$
nonzero DFT coefficients, out of $N=128$, and a number of available samples
$M=16$ within the transition region, when the recovery is not always obtained.
These realizations, when the recovery conditions, for a given signal and for
some of the considered sets of available samples, are met, can still be
detected. This process is especially important if we are not in the position
to define the sampling strategy for $M=N-Q$ available samples in advance, like
in the cases when the available samples are uncorrupted samples and their
positions can be arbitrary. The criterion for detection of a sparse signal in
recovery is the measure of the resulting signal sparsity. In this case
measures closer to the $\ell_{0}$-norm should be used for a detection. For
example, with $\ell_{1/4}$-form in the case of a false recovery all transform
coefficients are nonzero with $\sum_{k=0}^{N-1}\left\vert X(k)/N\right\vert
^{1/4}\sim N$. For a full recovery of a sparse signal the number of nonzero
coefficients (the measure value) is much lower since $s\ll N$.

Consider the case with $M=16$ available randomly positioned samples and $s=6$
nonzero DFT coefficients. Among 100 performed realizations a possible sparse
recovery event is detected (when the described sparsity measure of the result
is much lower than $N$). The DFT coefficients set of the detected sparse
signal is \ $\mathbb{K}_{s}=\{22$, $35$, $59$, $69$, $93$, $106\}$. It
confirms that the reconstructed signal is sparse. This sparse reconstruction
is checked for uniqueness using the theorem. The missing samples are from the
set $q_{m}\in\mathbb{N}_{Q}$. It is a set difference of all samples
$\mathbb{N=}\left\{  n\text{ \ }\left\vert 0\leq n\leq127\right.  \right\}  $
and
\[
\mathbb{N}_{A}=\left\{  7,\text{ }14,\text{ }18,\text{ }21,\text{ }34,\text{
}37,\text{ }51,\text{ }69,\text{ }79,\text{ }82,\text{ }89,\text{ }90,\text{
}99,\text{ }100\text{, }113,\text{ }117\right\}  .
\]
For $h=0,1,...,r-1=6$ corresponding values of $Q_{2^{h}}$ and $S_{2^{r-h}}$,
defined in the theorem, are calculated. Their values are:%

\[
\begin{tabular}
[c]{|rrrrrrrr|}\hline
$h$ & $0$ & $1$ & $2$ & $3$ & $4$ & $5$ & $6$\\\hline
$Q_{2^{h}}$ & $112$ & $58$ & $31$ & $16$ & $8$ & $4$ & $2$\\\hline
$S_{2^{7-h}}$ & $0$ & $0$ & $4$ & $5$ & $4$ & $4$ & $2$\\\hline
\end{tabular}
\]

\medskip

Note that $Q_{2^{0}}=112$ is the total number of missing samples, while
$Q_{2^{1}}$ is obtained by counting odd and even samples in $\mathbb{N}_{Q}$
and taking higher number of these two. Since there are $54$ samples at odd
positions and $58$ samples at even positions, it means that $Q_{2^{1}}=58$.

For $h=2$ there are $31$ missing sample $q_{m}\in\mathbb{N}_{Q}$ with
$\mod(q_{m},4)=0$, $26$ missing samples with $\mod(q_{m},4)=1$, $27$ missing
samples with $\mod(q_{m},4)=2$, and $28$ missing samples with $\mod(q_{m}%
,4)=3,$ resulting in $Q_{2^{2}}=\max\{31,26, 27,28\}=31$, and so on. We can
easily conclude that samples $x(1)$ and $x(65)$ are missing, meaning that
$Q_{64}$ assumes its maximal possible value $Q_{64}=2$.

Similar counting is done to get $S_{2^{7-h}}$. For example,
\[
S_{2^{7-6}}=S_{2^{1}}=\sum_{l=1}^{Q_{64}-1}P_{6}(l)=\sum_{l=1}^{1}%
P_{6}(l)=P_{6}(1)
\]
where array $P_{6}(l)$ is obtained by sorting number of even and odd elements
in $\mathbb{K}_{s}$. Since there are $2$ even and $4$ odd elements
$P_{6}(1)=2$ and $P_{6}(2)=4$ resulting in $S_{2^{1}}=2$.

\begin{figure}
[ptb]
\begin{center}
\includegraphics[
height=2.9033in,
width=3.1424in
]%
{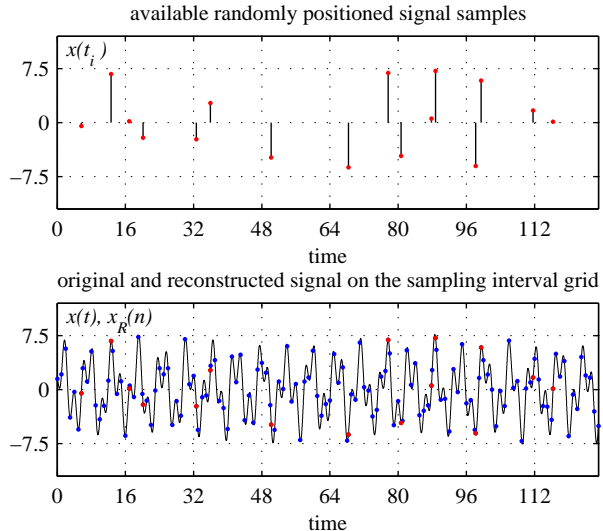}%
\caption{Available randomly positioned samples $x(t_{i})$ (red dots) of a
sparse signal $x(t)$ (top). Reconstructed signal $x_{R}(n)$ at the sampling
theorem positions (blue dots) along with the available samples (red dots)
(bottom). Continuous-time signal $x(t)$ is presented by solid line.}%
\label{grad_rand_samp_0_n_f_varijanta_lett_EVEN}%
\end{center}
\end{figure}

As expected this set of $112$ missing samples $\mathbb{N}_{Q}$ does not
guarantee a unique solution for an arbitrary signal of sparsity $s=6$. By
using the theorem with $S_{2^{r-h}}=0$ and $Q_{2^{h}}$ presented in the
previous table we easily get that the solution uniqueness for this set
$\mathbb{N}_{Q}$ and arbitrary signal requires $s<4$. However, for the
specific available signal values, a sparse signal is reconstructed in this
case, with nonzero coefficients at $\mathbb{K}_{s}=\{22$, $35$, $59$, $69$,
$93$, $106\}$. The uniqueness then means that starting from this signal we can
not find another signal of the same sparsity by varying the missing signal
samples positioned at $n\in\mathbb{N}_{Q}$. The theorem then gives the answer
that this specific recovered signal $x_{R}(n)$, with specific missing sample
values and positions $\mathbb{N}_{Q}$, is unique. It means that starting from
$x_{R}(n)$ we can not get another signal of the same or lower sparsity by
varying the missing samples only. The reconstructed signal is presented in
Fig.\ref{grad_rand_samp_0_n_f_varijanta_lett_EVEN}. The
signal-to-reconstruction-error ratio defined by (\ref{SRRR}), calculated for
all signal samples, is $SRR=111.08$ dB. It corresponds to the defined
reconstruction algorithm precision of about $100$ dB.

In addition to the considered case two obvious cases in the uniqueness
analysis may appear: 1) when both, the reconstructed signal and the worst case
analysis produce a unique solution using the set of missing samples
$\mathbb{N}_{Q}$, and 2) when both of them produce a result stating that a
signal with certain sparsity can not be reconstructed in a unique way with
$\mathbb{N}_{Q}$. Finally, it is interesting to mention that there exists a
fourth case when the set of missing samples can provide a unique
reconstruction of sparse signal (satisfying unique reconstruction condition if
it were possible to use $\ell_{0}$-norm in the minimization process), however
the $\ell_{1}$-norm based minimization does not satisfy the additional
restricted isometry property constraints \cite{RIP}, \cite{cosamp} to produce
this solution (the same solution as the one which would be produced by the
$\ell_{0}$-norm). This case will be detected in a correct way using the
presented theorem. It will indicate that a unique solution is possible using
$\mathbb{N}_{Q}$, while if the $\ell_{1}$-norm based minimization did not
produce this solution as a result of the reconstruction algorithm, the
specific reconstructed signal will not satisfy the uniqueness condition.%

\section{Conclusion}

Analysis of nonuniformly sampled sparse signals is performed. A gradient-based
algorithm with adaptive step is used for the reconstruction. A new criterion
for the parameter adaptation in the algorithm, based on the gradient
directions analysis, is proposed. It significantly improves the calculation
efficiency of the algorithm. \ The random nonuniformly positioned available
samples are recalculated based on the sampling theorem reconstruction formula.
Based on the new set of samples the recovery is performed. The methods are
checked and illustrated on numerical examples.

\end{document}